\documentclass[sn-mathphys]{sn-jnl}

\jyear{2023}%

\begin{document}

\title[Article Title]{Recommendation Algorithms: A Comparative Study in Movie Domain}


\author[1]{\fnm{Rohit } \sur{Chivukula}}\email{rohit.chivukula.kingdom@gmail.com}
\author*[1,2]{\fnm{T. Jaya Lakshmi}}\email{j.tangirala@shu.ac.uk}
\author[3]{\fnm{Hemlata} \sur{Sharma}}\email{h.sharma@shu.ac.uk}
\author[4]{\fnm{C.H.S.N.P. Sairam} \sur{Rallabandi}}\email{sairam.mtech20@gmail.com}

\affil[1]{\orgdiv{School of Computing and Engineering }, \orgname{University of Huddersfield}, \orgaddress{\city{Huddersfield}, \country{United Kingdom}}}

\affil[2]{\orgdiv{School of Engineering and Sciences}, \orgname{SRM University}, \orgaddress{ \state{Andhra Pradesh}, \country{India}}}

\affil*[3]{\orgdiv{School of Computing}, \orgname{Sheffield Hallam University}, \orgaddress{ \state{Sheffield}, \country{United Kingdom.}}}

\affil[4]{\orgdiv{Department of Computer Science and Engineering (Data Science)}, \orgname{R.V.R. \& J.C. College of Engineering}, \orgaddress{ \state{Andhra Pradesh}, \country{India.}}}

\abstract{
Intelligent recommendation systems have clearly increased the revenue of well-known e-commerce firms. Users receive product recommendations from recommendation systems. Cinematic recommendations are made to users by a movie recommendation system. There have been numerous approaches to the problem of recommendation in the literature. It is viewed as a regression task in this research. A regression model was built using novel properties extracted from the dataset and used as features in the model. For experimentation, the Netflix challenge dataset has been used. Video streaming service Netflix is a popular choice for many. Customers' prior viewing habits are taken into account when Netflix makes movie recommendations to them. An exploratory data analysis on the Netflix dataset was conducted to gain insights into user rating behaviour and movie characteristics. Various kinds of features, including aggregating, Matrix Factorization (MF) based, and user and movie similarity based, have been extracted in the subsequent stages. In addition to a feature in the XGBoost regression algorithm, the K-Nearest Neighbors and MF algorithms from Python's Surprise library are used for recommendations. Based on Root Mean Square Error (RMSE), MF-based algorithms have provided the best recommendations.
}

\maketitle

\section{Introduction}

Users are recommended products through recommender systems. Examples of items include movies, music, books, web sites, news, jokes, and restaurants. The job of recommendation utilises data such as user demographics, product features, and user activities such as buy/rate/watch. This historical data is collected in the form of ratings people have assigned to products. In certain instances, user behaviour such as songs listened to on music websites, movies viewed on cinema websites, and products purchased on e-commerce websites can provide insights. The adoption of sophisticated recommendation algorithms boosted Amazon’s profits by 35\%, BestBuy’s sales grew by 24\%, and Netflix and YouTube views each increased by 75\% and 60\%, respectively \cite{c1}. The importance of a personalised recommendation system extends far beyond the commercial arena, and includes fields such as health care, news, food, academia, and so on. Different aspects must be taken into account for various domains.

We emphasise on movie recommendation in this paper. The Movie recommendation task suggests movies to viewers based on their interest. Netflix is an online movie streaming app. Netflix recommends shows/Web-series/movies to the users. Netflix uses the ratings a user provides to already watched movies for this task. Netflix uses a system called CineMatch to accomplish this. CineMatch predicts whether a viewer enjoys a movie based on the rating given by him/her to the other movies. For better recommendation, Netflix has organized a competition on Kaggle.com. Netflix has released a massive amount of anonymous rating data. The winner of this competition wins an amount of \$1 million on improving recommendation accuracy by 10\%. Predicted movie ratings are measured in terms of how closely they line up with actual movie ratings. This challenge aims to (1) forecast a user's rating for an unrated movie and (2) reduce the difference between projected and actual ratings. The Netflix challenge ~\cite{c16} served as an inspiration for the research presented in this paper. 

\subsection{Problem Statement}
Given $(U,M,R)$ where $U$ is set of $m$ users given as $\lbrace u_1, u_2, \ldots,u_m \rbrace$, $M$ is a movie set of $n$ movies given as $M=\lbrace m_1, m_2, \ldots,m_n \rbrace$ and $R$ denoting ratings matrix of size $m X n$. $R_{ij}$ denotes the rating given by user $u_i$ to movie $m_j$. The aim of recommender system is to suggest a list of movies to   users that the users did not watch/rate \cite{c4}. This is done by predicting the unfilled entries of the rating matrix $R$. From machine learning perspective, this task is reduced to minimize the discrepancy between the actual and predicted ratings. Thus, the recommendation task is an optimization problem, whose objective function is given in Eq.\ref{eq1}.

\begin{equation}
	\label{eq1}
	Min \sum_{i,j}\|R_{ij} - R'_{ij}\|^2
\end{equation}

Where $R_{ij}$ and $R'_{ij}$ are the actual and predicted ratings respectively.

\section{Related Work}

The literature on recommender systems is expanding rapidly, with new algorithms being proposed seemingly daily. Recommender system solutions include content-based recommendation, Collaborative Filtering (CF), Hybrid Approach, and regression models. These methods are briefly described below.

\subsection{Content-based recommendation (CBR)}
\begin{figure}[h]
    \centering
    \includegraphics[width=1\textwidth]{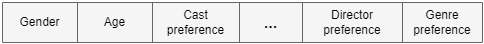}
    \caption{Sample user profile}
    \label{user_profile}
\end{figure}

$CBR$ begins with the creation of profiles for both the user and the item being recommended (in this case, a movie) ~\cite{c9}. An example user profile is given in Fig. \ref{user_profile}. 

$u_i$ can be recommended with products based on user-user similarities as follows: Using any measure of similarity between users' profiles, first determine how similar $u_i$ and all other users are. The notion of cosine similarity is well-known in this field. Afterwards, select the most comparable users of $u i$. Recommend movies that they have enjoyed or rated high to $u_i$

\begin{figure}[h]
    \centering
    \includegraphics{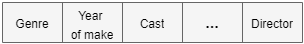}
    \caption{Sample movie profile}
    \label{movie_profile}
\end{figure}

A movie profile can be built in a similar fashion. A sample movie profile is shown in Fig. \ref{movie_profile}.

The similarity between two movies can be used to recommend a movie to a user $u_i$ as follows: Let $u_i$ has rated $m_j$. Calculate the similarity between $m_j$ and other movie-vectors to see how similar they are. Choose the best $n$ movies that are identical to $m_j$ and suggest them to $u_i$.
The following are some advantages of $movie-to-movie$ similarity over $user-to-user$ similarity.
\begin{itemize}
    \item Users' tastes change over time, which might lead to the creation of new user profiles.
    \item In comparison to the quantity of movies available, the number of users is astronomical. Because of this, calculating the similarity between $movie-movie$ and the similarity between $user-user$ is less.
\end{itemize}

Unrated items can be recommended by $content-based$ recommendation algorithms. It is possible to explain CBRs' recommendations because they are based on the item's features \cite{c2}. CBRs necessitate in-depth information on the users and the items being tracked. For items that cannot be clearly distinguished, the recommender fails.

\subsection{Collaborative Filtering $(CF)$}
User-specific recommendations are made possible by collaborative filtering approaches. These methods are based solely on past behaviour and do not require any more input from the user ~\cite{c7},~\cite{c10}. CF approaches use a rating matrix as input, with rows and columns corresponding to users and items, and cells that represent the user's ratings for each item in the matrix. The collaborative filtering approach works like this: It is more likely that $u_1$ will agree with $u_2$ on another topic than it is that $u_1$ will agree with a random user \cite{c11} if their judgements on the same problem are identical. Model-based CF and memory-based CF are the two main types of CF approaches.

\begin{figure}[h]
	\includegraphics[width=1\textwidth]{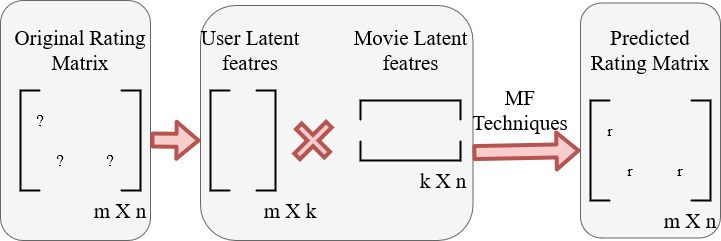}
	\caption{Matrix Factorization}
	\label{MF}
\end{figure}

\textbf{Model-based methods} Matrix factorization algorithms are most popular in this category. These methods extract hidden(latent) features of users and items from by factorizing the rating matrix. The product of a user's latent vector and that of an item yields the user's rating for the item. Fig \ref{MF} describes the process.

Data sparsity is a severe issue with CF algorithms. Dimensionality reduction approaches, Principal Component Analysis (PCA), and Singular Value Decomposition (SVD) are among popular sparsity-resolution techniques. However, when particular individuals or products are removed, essential information for associated suggestions may be lost, and recommendation quality may suffer ~\cite{c3}. The cold start problem is another shortcoming of collaborative filtering systems. As there is no historical purchasing history for a newly introduced item, these algorithms cannot recommend them to existing users. 

\textbf{Memory based methods} employ Pearson-coefficient, cosine/Manhattan distance to compute nearest neighbours for users and items. It is possible to further subdivide memory-based CF approaches into two types: user-based and item-based~\cite{c14}. Users' ratings on items are used to compute the degree of similarity between them \cite{c8}. Unrated items of a user are assigned to a group of nearest neighbours, and the ratings of those neighbours are used to anticipate the user's rating for those items. Item neighbourhood shows the number of users who have rated the same things \cite{c5}. It is thus possible to estimate a user's item rating based on the other users in their user neighbourhood and the other items in their user neighbourhood. According to \cite{c22}, the rating matrix is modelled as a graph and the similarity is calculated based on the neighbourhood.

\textbf{Hybrid approach} incorporates both information of collaborative and content based. Content boosted CF algorithm ~\cite{c6}, makes advantage of the user profile data to suggest items to new users.

\subsection{Regression}
In machine-learning, regression is defined as follows: Given $m$ pairs $\lbrace(F_1,t_1)$, $(F_2,t_2) \ldots (F_m, t_m) \rbrace $, each $F_i \in R^n$ describing a feature vector describing $i^{th}$ data point and $t_i \in R$ is the corresponding target value, the regression task fits a mapping that predicts the value of $l$ for a given $F$. 

\begin{figure}[h]
	\includegraphics[width=1\textwidth]{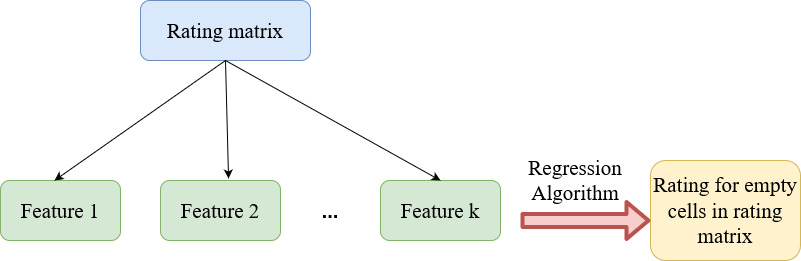}
	\caption{Recommendation modelled as regression}
	\label{regression}
\end{figure}

As ratings of rating matrix will be commonly numerical, the prediction of ratings can be modeled as a regression task. From the rating matrix, the features needed by the regression technique can be retrieved using traditional similarity metrics applied to the relationships between users and items. The process is shown in Fig.\ref{regression}.

Regression models and collaborative filtering techniques are used to solve the problem of movie recommendation in this study. At \cite{c16}, the Netflix dataset was made available for experimentation. The dataset and the exploratory data analysis are detailed in the following section.

\section{Dataset and Insights}

There are approximately 100M ratings from 480K Netflix customers in the dataset published by Netflix in the Kaggle \cite{c16}. The information between October 1998 and December 2005 was gathered. Rating ranges from 1 to 5. The identities of the customers have been masked. For each movie, the date it was rated, as well as the title and year were released. Table.\ref{dataset_description} contains a list of the files included in the dataset.

\begin{table}[h]
	\caption{Dataset Description}\label{dataset_description}
	\begin{tabular}{lll}
		\hline
		Name of the file  &  Description & Format \\ \hline
		training\_set.rar & \begin{tabular}[c]{@{}l@{}} compressed directory \\  consisting of 17,770 individual files, \\ one for each movie.\end{tabular}        
  
         & \begin{tabular}[c]{@{}l@{}}MovieID:\\ CustomerID, Rating, Date\end{tabular}                                                           \\ 		
		movie\_titles.txt & \begin{tabular}[c]{@{}l@{}}Describes the  details \\ about movie.\end{tabular}                                     & MovieID, YearOfRelease, Title  \\ 
		qualifying.txt    & 
        \begin{tabular}[c]{@{}l@{}} The proposed algorithms should \\  predict the  user ratings to \\ movies in qualifying dataset\end{tabular} & MovieID: CustomerID, Date \\ 
		probe.txt         & \begin{tabular}[c]{@{}l@{}}Provided to be   used before \\ evaluating the prediction. \\ Consists of a list of customers \\ that rated a movie.\end{tabular}                                                         & \begin{tabular}[c]{@{}l@{}}MovieID:\\ CustomerID1\\ CustomerID2\\ \ldots\end{tabular}                                                         \\ \hline
	\end{tabular}
\end{table}

\subsection{Train and Test set}
An 80\% training set and a 20\% percent testing set are extracted from the Netflix movie dataset to be used for training and testing purposes. Accordingly, the training set contains ratings from earlier years, while the test set contains ratings from more recent years. Table.\ref{dataset_statistics} shows the breakdown of ratings, users, and movies into training test segments for each of the various types of tests.

\begin{table}
	\begin{center}
	\caption{Dataset Statistics}
	\label{dataset_statistics}
	\begin{tabular}{llll}
		\hline
		\textbf{Dataset} & \textbf{\#Ratings} & \textbf{\#Users} & \textbf{\#Movies} \\ \hline
		Total            & 100,480,507        & 480,189          & 17,770            \\ 
		Training         & 80,384,405         & 406,041          & 17,424            \\ 
		Testing          & 20,096,102         & 349,312          & 17,757            \\ \hline
	\end{tabular}
\end{center}
\end{table}

\subsection{Exploratory Data Analysis on Training Data}
\begin{figure}
	\centering
	\includegraphics[width=0.7\textwidth]{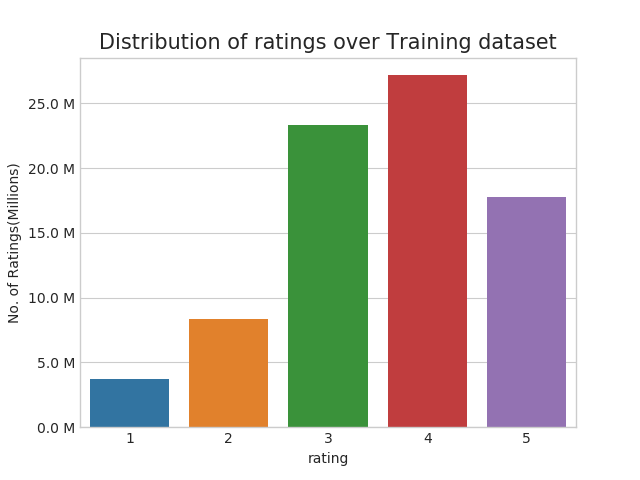}
	\caption{Rating's distribution}
	\label{Distribution_of_ratings_over_training_set}
\end{figure}

Experimentation with data is greatly aided by exploratory data analysis (EDA). This section details the results of EDA on the training data. Fig.\ref{Distribution_of_ratings_over_training_set} depicts the rating distribution that was used to better analyse the data.

\begin{figure}
	\centering
	\includegraphics[width=1\textwidth]{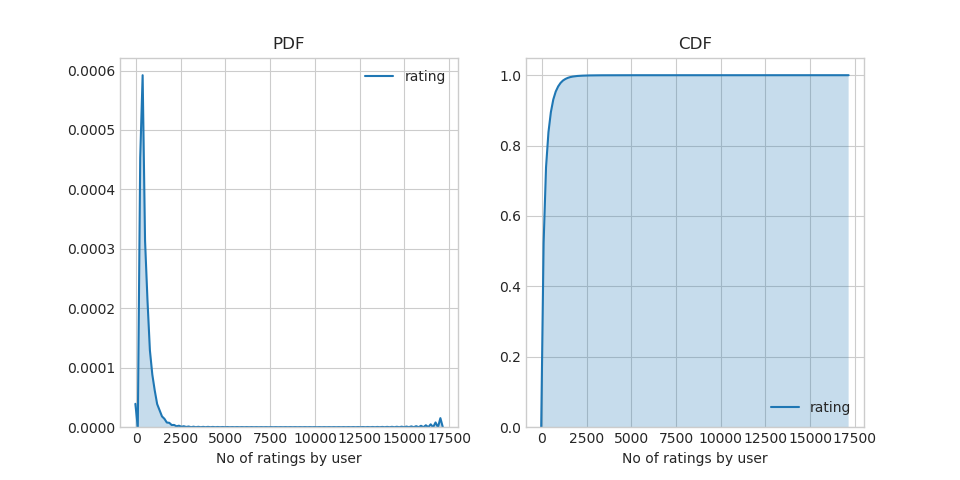}
	\caption{Ratings Analysis}
	\label{Analysis_on_the_Ratings_given_by_user}
\end{figure}

From EDA, it is evident that users tend to assign a rating of 4 the most frequently. Figure. \ref{Analysis_on_the_Ratings_given_by_user}, displays the results of a breakdown of the ratings given by each individual user. Couple of people have given the app extremely high ratings, and the majority of users have given a few stars at most.

\begin{figure}
	\centering
	\includegraphics[width=0.7\textwidth]{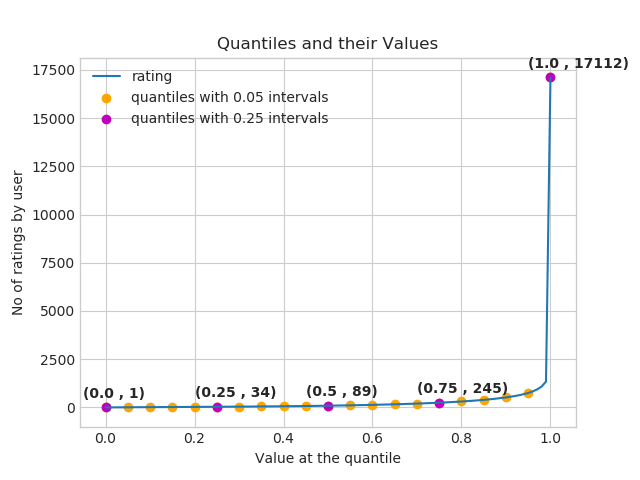}
	\caption{Quantile analysis of user ratings}
	\label{Quantiles_analysis_of_the_ratings_given_by_a_user}
\end{figure}

The average number of ratings by a user is 198, according to the quantile analysis plot presented in Fig. \ref{Quantiles_analysis_of_the_ratings_given_by_a_user}. The minimum user rating is found to be one, max user ratings 17,112, and 50\% of users rated 89 movies. 

\begin{figure}
	\centering
	\includegraphics[width=0.7\textwidth]{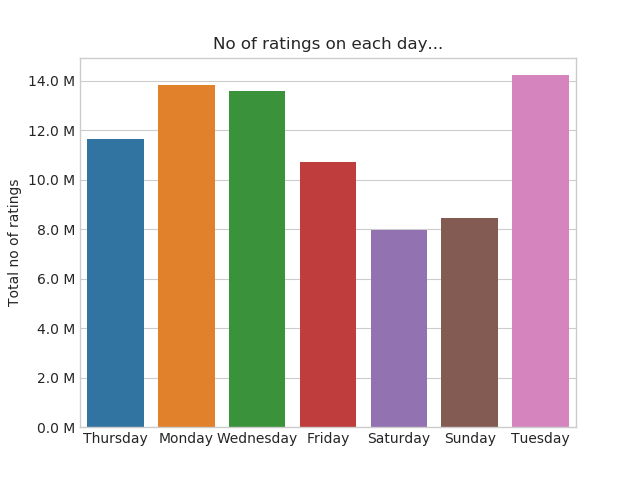}
	\caption{Day-wise rating analysis}
	\label{Day_wise_analysis_of_ratings}
\end{figure}
 
\begin{figure}
	\centering
	\includegraphics[width=1\textwidth]{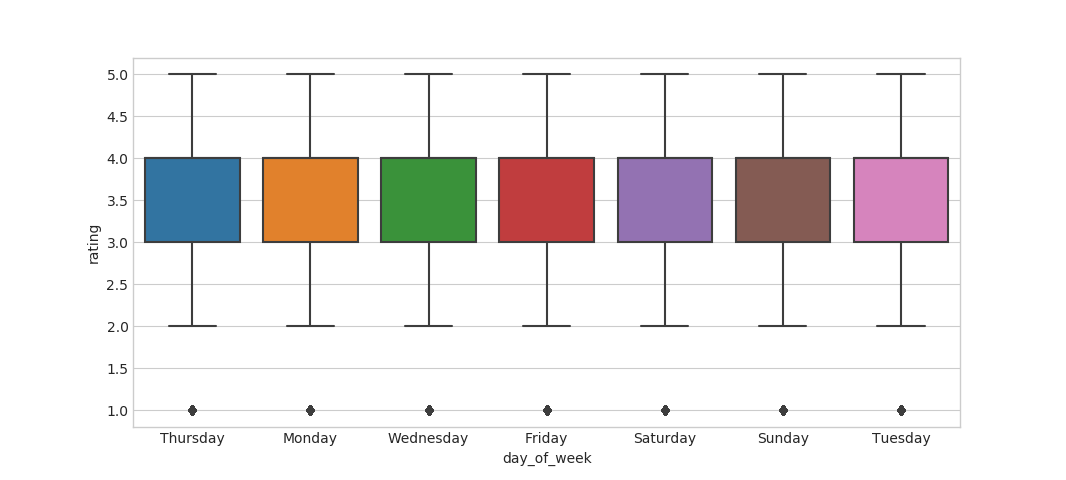}
	\caption{Box-plot depicting rating on week-day}
	\label{Box_plot_showing_weekday_Vs_rating}
\end{figure}

From Fig.\ref{Day_wise_analysis_of_ratings}, it is evident that the number of ratings is dropped during weekend.  The boxplot shown in Fig.\ref{Box_plot_showing_weekday_Vs_rating} shows that the day of the week does not appear to have a significant impact on the rating, as the plot is steady on all days. As a result, the day of the week is no longer considered a likely factor in determining a movie's rating.

Collaborative filtering is plagued by sparsity. The sparsity of given train matrix is 99.82 and that of the test matrix is 99.9. This demonstrates that around 99\% of train and test matrix entries are empty. 

\section{Proposed Approach to Predict Ratings}
In this work, we use three kinds of features. 
\begin{enumerate}
	\item Aggregate features 
	\item User similarity features and 
	\item Movie similarity features. 	
\end{enumerate}

\begin{figure}[h]
	\centering
	\includegraphics[width=0.7\textwidth]{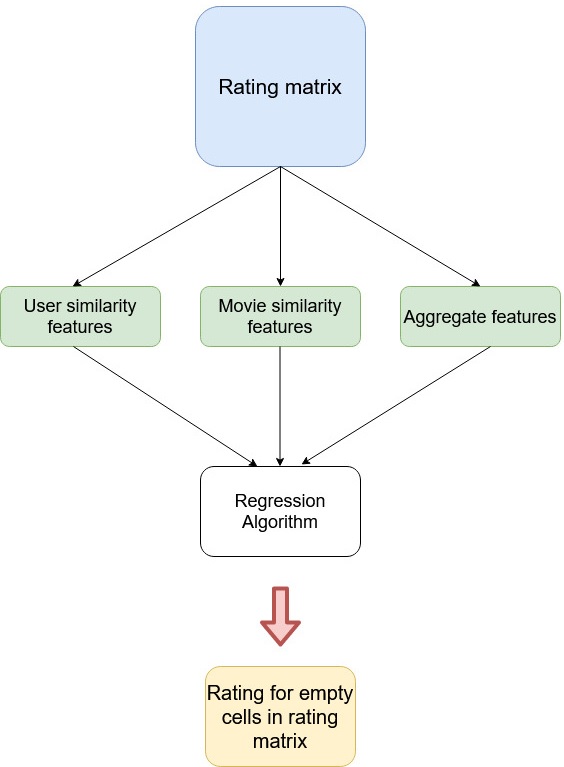}
	\caption{Approach for movie recommendation}
	\label{top_level_approach}
\end{figure}

All these features are incorporated into several regression models in order to forecast the final score. The top-level diagram depicting the strategy taken in this study is shown in Fig.\ref{top_level_approach}. Feature computations are explained in the following sections. In all the calculations, $m$ represents \#users and $n$ is \#movies.
    
\subsection{Computation of Aggregate features}
The list of aggregate features is given below.
\begin{itemize}
	\item \textbf{Global average (GAvg)}: Average user rating for all $n$ movies. 
	\item \textbf{User average (UAvg)}: This is a $m$ element vector, where $i^{th}$ element in the vector represents the average rating given by user $u_i$. A user's usual rating behaviors is depicted in this vector. There are some individuals who consistently give negative ratings to movies, while some rate a lot of movies highly. 
	\item \textbf{Movie average (MAvg)}: In this $n$-element vector, each $i^{th}$-element reflects the average rating of movie $m_i$. This shows whether a film is a hit or a total bust. For instance, the movie Titanic has been well received by audiences of all ages. The film was a massive success and has been given good ratings by many users no matter when they first saw it.
\end{itemize} 

It is necessary to develop similarity matrices in order to compute the user and movie similarity features. The rating of a movie given by user is documented in the dataset. This user-item matrix has many empty cells, because Netflix users can't rate every movie. Similarly, a movie wont be viewed by every user. Using a user-user similarity matrix, it is possible to determine how similar two users are. Similarly, a film-film similarity matrix can be used to identify comparable films. 

\subsection{Computing user-user similarity}
\label{user-user-matrix}
A user-user similarity matrix is of size $n X n$ where $n$ is the number of users. A user $i$ is represented by the matrix's $i^{th}$ row. The matrix entries are as in Eq. \ref{eq2}.
\begin{equation}
\label{eq2}
	U_{ij} = The~similarity~score~of~user~i~to~user~j.
\end{equation}

The size of the matrix for Netflix dataset has 405K rows and columns, which makes this computation time demanding. The training set has 405,041 participants. The dimensions of each user vector are 17K. Calculating similar users for a single user took the implementation an average of 8.88 seconds. The similarity matrix can be computed in 41.6 days by performing a simple calculation. The computation is shown below.

\begin{equation*} 
	\begin{split}
	405041 * 8.88 &= 3596764.08~sec \\
	&=59946.068~min \\
	&=999.101133333~ hours \\
	&=41.629213889~ days \ldots
	\end{split}
\end{equation*}

Therefore, the influence of dimension reduction techniques on the speed of the process is examined. 500-dimensional vectors are computed using SVD. On a computer with an i5 processor and 16GB of RAM, computing similar users for a single user took approximately 12.18 days. The computation time with 405,041 users in the training set is 57 days.


The denseness of the reduced vector is the cause for the high computation time. Because of this, it can be concluded that the dimensionality reduction technique did not aid in decreasing computation of the similarity matrix. Therefore, a user's similarity to other users is calculated as necessary.

\subsection{Computing movie-movie similarity}
Similarly, a movie-movie matrix of size $m X m$ can be computed, where $m$ is the \#movies. There are a variety of ways to compute the similarity between two vectors in the literature. We use Cosine similarity matrix. The cosine similarity on a multidimensional plot captures the orientation (angle) of the data objects but not the quantity. The closer the angle, the greater the resemblance between the data points(movies). Movie vectors are 405,000 dimensional and sparse. Though the vectors are of high dimensional, matrix computation isn't a problem because of sparseness. Moreover, the similarity of a movie with all other movies is not needed. 

\begin{figure}[h]
	\centering
	\includegraphics[width=0.7\textwidth]{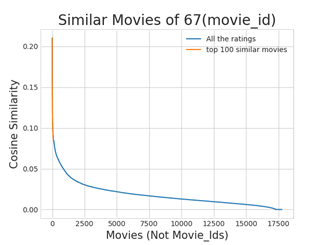}
	\caption{Similarity of movie-Id:67}
	\label{Similarity_of_movie_Id_67}
\end{figure}

"Vampire-Journals" movie with Id 67 has been selected to demonstrate how movie-movie similarity works. 270 Users have given it a rating. It is similar to 17284 other movies. Fig.\ref{Similarity_of_movie_Id_67} depicts the similarity of the movie with id 67.

It is evident from Fig.\ref{Similarity_of_movie_Id_67} that top 100 movies are similar with a threshold 0.08. 
\subsection{Feature Vector}
\begin{figure}[h]
	\centering
	\includegraphics[width=1\textwidth]{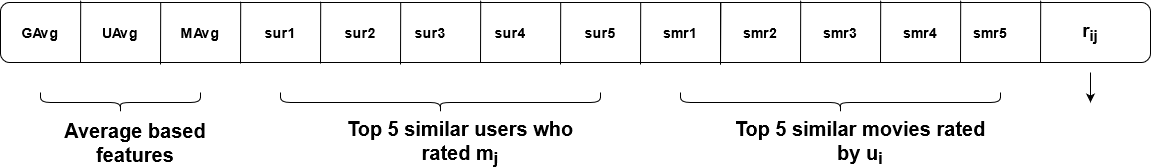}
	\caption{Feature vector of $(u_i,m_j)$}
	\label{Feature_vector}
\end{figure}

To predict the rating $r_{ij}$, given by user $u_i$ to a movie $m_j$, a feature vector corresponding to user $u_i$ and movie $m_j$ is represented in Fig. \ref{Feature_vector}, which consist of the following features.
\begin{itemize}
	\item Aggregate based features
	\begin{itemize}
	\item G\_Avg: Aggregated average rating
	\item U\_Avg: Average of ratings given by user $u_i$
	\item M\_Avg: Average ratings given to $m_j$
	\end{itemize}
	\item Similarity values of top 5 similar users that rated $m_j$ 
	\item Similarity values of top 5 similar movies rated by $u_i$. 	
\end{itemize}

So, feature vector of$ (u_i,m_j) = (GAvg, UAvg, MAvg, susr_1,susr_2,susr_3, susr_4,susr_5,smvr_1,smvr_2,smvr_3,smvr_4,smvr_5)$
where $susr_i$ represents Similarity value of the top $ i^{th}$ similar user that has rated movie $m_j$  and 
and $smvr_i$ denotes similarity value of the top $i^{th}$  similar movie that is rated by user $u_i$. 	

\section{Implementation and Results}
\subsection{Machine Learning models}
A regression model is used to solve the movie recommendation problem.We employ  XGBoost ~\cite{c18}, Suprise Baseline Model ~\cite{c19}, Surprise KNNBaseline ~\cite{c20} and their combinations in this work. TFollowing is a quick explanation of each of these methods.
\begin{itemize}
	\item XGBoost is an ensemble algorithm that uses a gradient boosting framework in decision tree ~\cite{c23}. The parallel tree boosting method provided by XGBoost is often both quick and accurate. Each prediction tree's performance is averaged out in ensemble learning, which combines predictions from several different models. Boosting is a strategy for correcting mistakes made by prior models by increasing the weights of the models. XGBoost has the ability to handle big datasets surpassing RAM extremely effectively. A variety of regularizations are supported by this approach, which helps reduce overfitting.It has the ability to prune a tree, manage missing values, and deal with outliers to some extent. The prediction power of XGBoost changes with features used in that framework. 
	\item Surprise Baseline Model ~\cite{c19}: This model uses Equation.\ref{eq5} to predict ratings.
	\begin{equation}
		{r'}_{ij}=b_{ij}=\mu+b_i+b_j  \label{eq5}
	\end{equation}
	
	where, $\mu$ is average of all ranks in the training data, $b_i$ is user bias and $b_j$ is movie bias.
	\item Surprise KNNBaseline with user-user similarities is computed using  Eq.\ref{eq3}.
	\begin{equation}
	{r'}_{ij}=b_{ij}+\frac{\displaystyle \sum_{u \in N(i,j,k)}sim(i,u).(r_{uj} - b_{uj})}{\displaystyle \sum_{u \in N(i,j,k)} {sim(i,u)}} \label{eq3}
	\end{equation}
	
	Where, $b_{ij}$ is baseline prediction of (user $i$, movie $j$) rating, $N(i,j,k)$ is the set of $k$ similar users of user $i$, who rated movie $j$, $sim(i,u)$ is the similarity between users $i$ and $u$ computed using shrunk Pearson-baseline correlation coefficient, which is based on Pearson baseline similarity. We consider baseline predictions instead of the mean rating of user/movie. 
	\item Surprise KNNBaseline ~\cite{c20} with movie-movie similarities is computed similar to Equation. \ref{eq3}, by replacing users by movies.
\end{itemize}

Along with these regression models, the other combinations worked out are given below.
\begin{enumerate}
	\item   XGBoost+Surprise Baseline predictor+KNNBaseline predictor.
	\item	SVD MF on User-Movie matrix \cite{c21}.
	\item	SVD MF with user given implicit feedback.
	\item	XgBoost+Surprise Baseline+Surprise KNNbaseline+MF 
	\item	XgBoost with Surprise Baseline+Surprise KNNbaseline+MF	
\end{enumerate}

\section{Performance Metrics}
Having a single numerical to evaluate model's performance is tremendously advantageous. 
Root Mean Square Error (RMSE) is a commonly used metric for assessing the accuracy of regression models. Mean Absolute Percentage Error (MAPE) was also utilised as a performance statistic in this study, along with RMSE. Both of these metrics are defined as follows:

\subsection{Root Mean Square Error \cite{c12}}
Mean Square Error (RMSE) uses the mean value of all squared variations between expected and actual ratings and computes square root of it. The mathematical definition of RMSE is given in Equation.\ref{eq_rmse}

 \begin{equation}
\mathrm{RMSE} = \sqrt{\frac{\sum_{i,j} \left( r_{ij} - r^{\prime}_{ij} \right)^2}{n}}
\label{eq_rmse}
\end{equation}

Where $r_{ij}$ is the rating that user $u_i$ has assigned to movie $m_j$ and $r'_{ij}$  is the predicted rating. The ideal value of RMSE is 0, denoting that the predicted ratings are close to actual. The value of RMSE depends on the data and the range of ratings. Bigger errors are penalised by the RMSE. The more accurate the model is, the lower the RMSE.

\subsection{Mean Absolute Percentage Error ~\cite{c15}}
Another way to gauge the accuracy of forecasts is to look at the mean absolute percentage error (MAPE). This measure computes accuracy as a percentage.  MAPE is computed using Eq.\ref{eq_mape}.

\begin{equation}
MAPE=\frac{1}{n}\sum_{i,j}\frac{r_{ij} - r'_{ij}}{r_{ij}}*100  \label{eq_mape}
\end{equation}

Where $r_{ij}$ and $r'_{ij}$ are exactly same as in Equation.\ref{eq_rmse}.
Ideally, MAPE should be zero. MAPE is simple to interpret. A MAPE of 10 indicates a 10\% discrepancy between predicted and actual scores. MAPE may be infinite if the actual rating is 0 for some items/products. MAPE may provide extraordinarily large percentage mistakes if the real values are exceedingly small and near to zero. 

These selection of these measures depends on the data. In the next section, an exploratory data analysis(EDA) is performed. Using exploratory data analysis(EDA), machine learning algorithms can identify the strong and weak features from data. 
\section{Results and Discussion}
\begin{figure}
    \centering
    \includegraphics[width=1\textwidth]{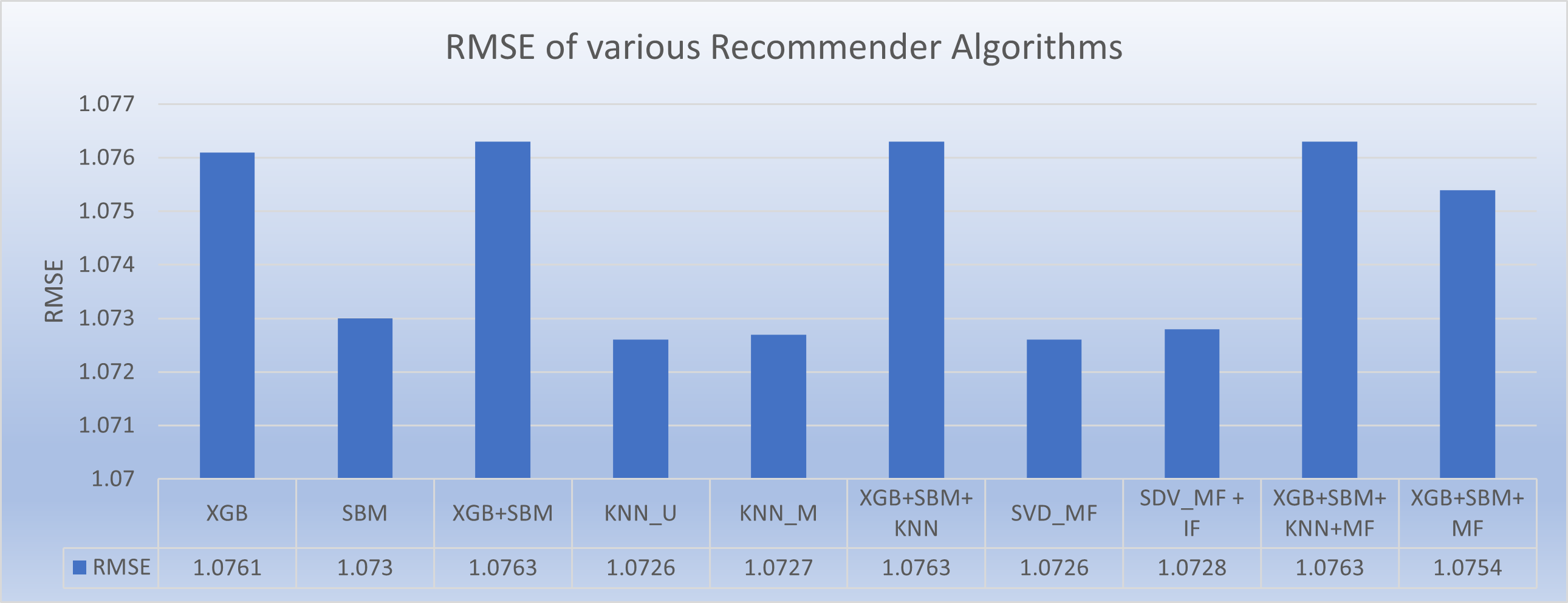}
    \includegraphics[width=1\textwidth]{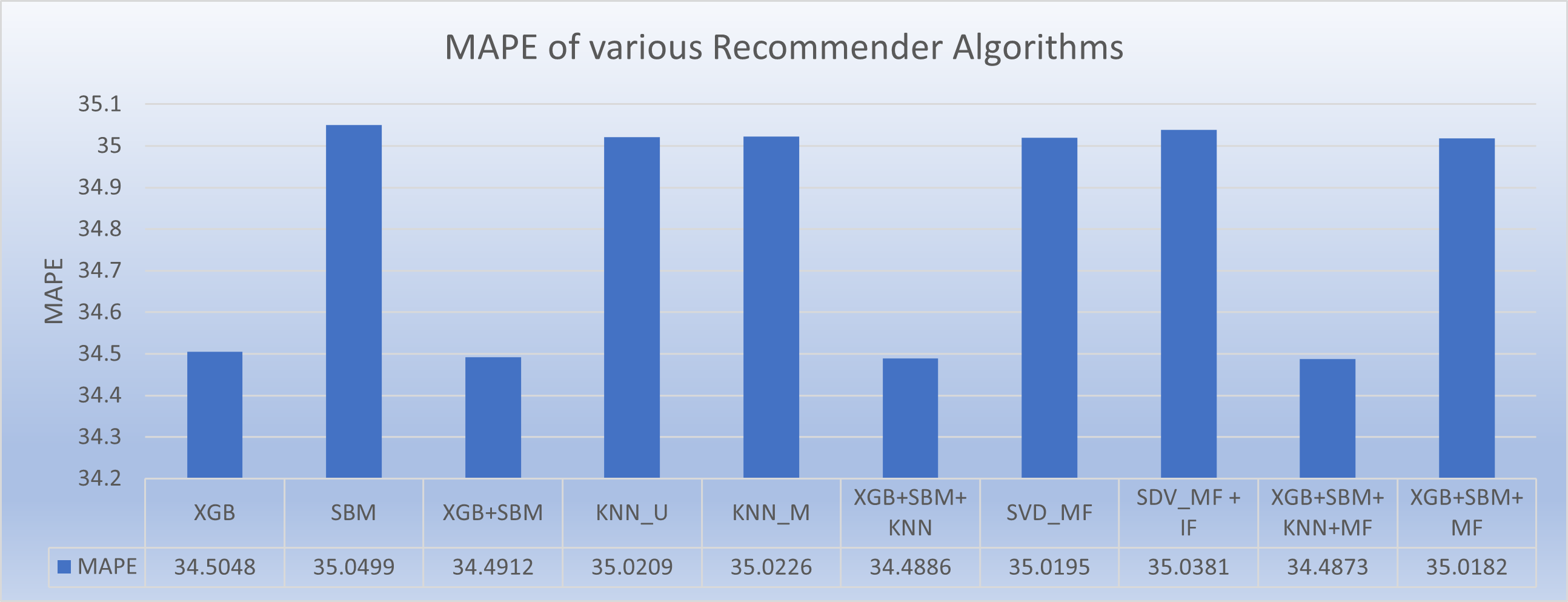}
    \caption{Performance of various Recommender Algorithms}
    \label{results_fig}
\end{figure}

Figure.\ref{results_fig} shows the outcomes of the algorithms. 

There is just a small variation in MAPE scores, ranging from 34\% to 35\%. RMSE of these algorithms vary by 33\%. Results show that SVD-based models and Surprise KNNBaseline with user-user similarities achieve the lowest(best) RMSE, while XGBoost with 13 features + Surprise baseline predictor and XgBoost + Surprise Baseline + Surprise KNNbaseline + MF Techniques provide the highest(least performance) RMSE. The use of extra features appears to have a detrimental effect on performance.


The train and test data have been split based on temporal information. It's possible that this separation will cause some issues. For example, some users may not be in the train set but in the test set since they recently joined. Some movies that were released during the testing period may not be available during the training time. For these users and movies, there is no rating data. This is referred to as a "cold start." There's a chance that this will have an impact on our recommendation. Users who didn't rate throughout training period and movies that didn't get any ratings were analysed in the dataset to get an idea of how cold start affects users and movies. 

It is observed that Out of 480189 users, only 405041 users are present in the training data, indicating 75148(15.65\%) users did not appear in training data. Predicting their ratings is difficult. Similarly, out of 17770 movies present in the dataset, train data contains only 17424. This means there are 346(1.95\%) movies, which do not appear in the training set, leading to a cold start problem. This suggests that users are more affected by the cold start issue than movies.

Only 405,041 out of a total of 480,189 users are present in the training set indicating that 75,148 (15.65\%) users were new in test data. Int he same way, out of a total of 17,770 movies in the dataset, only 17,424 are included in the training set. This results in a cold start issue. This indicates that users are more impacted by the cold start problem than movies.

\section{Conclusion}
Movie recommendation problem is critically investigated in this work. The problem is addressed using regression as well as matrix factorization problem. The features for regression model use the features such as global rating average, user rating average, and movie rating average and the output of a few algorithms from the surprise library. The experiments use Kaggle.com's Netflix prize challenge. MF methods such as SVD outperformed all other models. Our future work focuses on more novel methods to improve the performance of movie recommendation.

\paragraph{Declaration}

\begin{itemize}
    \item Competing Interests - The authors do not have any conflict of interest.
    \item Funding Information - Not Applicable
    \item Author contribution -The concept was conceived by Jaya Lakshmi, who also handled the documentation. Rohit Chivukula was responsible for the design and implementation, while Hemlata Sharma conducted the Exploratory Data Analysis. Sairam contributed by analyzing the results.
    \item Data Availability Statement - Not Applicable
    \item Research Involving Human and /or Animals - Not Applicable
    \item Informed Consent - Not Applicable
\end{itemize}


\bibliography{MovieRS}%
\end{document}